# Generalized Lorentz reciprocal theorem in complex fluids and in non-isothermal systems

Xinpeng Xu[1], Tiezheng Qian[2,*]

[1]Physics Program, Guangdong Technion—Israel Institute of Technology,
Shantou, Guangdong 515063, People's Republic of China

[2]Department of Mathematics, Hong Kong University of Science and Technology,
Clear Water Bay, Kowloon, Hong Kong

The classical Lorentz reciprocal theorem (LRT) was originally derived for slow viscous flows of incompressible Newtonian fluids under the isothermal condition. In the present work, we extend the LRT from simple to complex fluids with open or moving boundaries that maintain non-equilibrium stationary states. In complex fluids, the hydrodynamic flow is coupled with the evolution of internal degrees of freedom such as the solute concentration in two-phase binary fluids and the spin in micropolar fluids. The dynamics of complex fluids can be described by local conservation laws supplemented with local constitutive equations satisfying Onsager's reciprocal relations (ORR). We consider systems in quasi-stationary states close to equilibrium, controlled by the boundary variables whose evolution is much slower than the relaxation in the system. For these quasi-stationary states, we derive the generalized Lorentz reciprocal theorem (GLRT) and global Onsager's reciprocal relations (GORR) for the slow variables at boundaries. This establishes the connection between ORR for local constitutive equations and GORR for constitutive equations at boundaries. Finally, we show that the LRT can be further extended to non-isothermal systems by considering as an example the thermal conduction in solids and still fluids.

## I. INTRODUCTION

H. A. Lorentz derived in 1896 [1] a reciprocal theorem governing the slow viscous flows of incompressible Newtonian fluids under the isothermal condition. The classical Lorentz reciprocal theorem (LRT) has subsequently found wide applications [2-4], especially to flows at low Reynolds number in fluid/particle composite systems, such as suspensions, emulsions, and porous media. The LRT can be derived [5] from the fundamental thermodynamic reciprocal relations formulated by L. Onsager in 1931 [6,7] for linear irreversible processes. Onsager's reciprocal relations (ORR) are valid for the linear response of fluxes to forces in the vicinity of thermodynamic equilibrium [8-17]. These relations assert that due to the microscopic time-reversal symmetry, the coefficient matrix that couples the forces and fluxes must be symmetric. Note that to apply ORR, a set of conjugate forces and fluxes must be identified properly first, with the sum of their products equal to the rate of entropy production [8-17]. Based on the reciprocal symmetry, Onsager [6,7] also formulated a variational principle that is typically used to derive local constitutive equations. Onsager's variational principle has opened up a straightforward and unified way of deriving dynamic equations for complex fluids and soft matter [18-26]. It is worth pointing out that although the validity of ORR requires the near-equilibrium condition, it does not mean that ORR cannot be used to study thermodynamic processes far from equilibrium [8-26]. Based on the hypothesis of fast local equilibration of small mass elements, the near-equilibrium condition is realized locally, and hence ORR can be validated and employed [8-10].

*Corresponding author: maqian@ust.hk

In the present work, we generalize the LRT from simple fluids to complex fluids in which the hydrodynamic flow is coupled with the evolution of internal degrees of freedom [9-14]. We derive the generalized Lorentz reciprocal theorem (GLRT) for two typical complex fluids: two-phase binary fluids [22, 23] and micropolar fluids [10, 11], in which the solute concentration and spin act as internal degrees of freedom, respectively. Technically, we first use Onsager's variational principle to derive the system of dynamic equations for local variables. We then construct the free energy balance equation for non-equilibrium quasi-stationary states that are maintained by open or moving boundaries. We finally derive the GLRT for the quasi-stationary states in the close proximity of equilibrium state. Physically, the GLRT is derived under the following conditions: (i) the flow is so slow that the inertial effect is negligible, (ii) the reciprocal symmetry in local constitutive equations is preserved, and (iii) the quasi-stationary states are in the close proximity of equilibrium state such that the dynamic system can be linearized. To demonstrate the applications of the GLRT, we consider a few special system geometries in which a set of conjugate forces and fluxes are identified at the system boundaries and global Onsager's reciprocal relations (GORR) are obtained for the coefficient matrix that couples these boundary forces and fluxes. Finally, we show that the GLRT and GORR can be further generalized to non-isothermal systems by considering as an example the thermal conduction in solids and still fluids.

## II. TWO-PHASE BINARY FLUIDS ON SOLID SURFACES

### A. Hydrodynamic equations derived from Onsager's variational principle

Consider a two-phase binary fluid flowing on rigid solid surfaces [22,23]. For simplicity, we proceed under the following assumptions [18,23]. (i) The two fluid components have identical molecular volume and identical molecular mass before mixing. (ii) The volumes of the two components are additive after mixing. Such a simple binary fluid then has the following properties. (i) The mass fraction of each component equals to their volume fraction, and hereafter the volume fraction of the solute is denoted by $\phi$. (ii) The mass-averaged velocity equals to the volume-averaged velocity, hereafter denoted by $\mathbf{v}$. (iii) The binary fluid is incompressible with a constant mass density independent of space and time.

For a binary fluid with two-phase coexistence, its thermodynamic properties can be uniquely defined by a Ginzberg-Landau-type free energy functional [22,23]

$$F[\phi(\mathbf{r})] = \int \left[ f(\phi) + \frac{K}{2}(\nabla \phi)^2 \right] d\mathbf{r}, \tag{1}$$

in which $f(\phi)$ is a single-phase free energy density that describes the phase behavior of the binary fluid, and $K$ is a parameter associated with the fluid-fluid interfacial thickness and interfacial tension. Minimizing the free energy functional $F[\phi(\mathbf{r})]$ with respect to $\phi(\mathbf{r})$ gives the equilibrium conditions

$$\mu = \frac{\delta F}{\delta \phi} = \frac{df(\phi)}{d\phi} - K\nabla^2 \phi = \text{const.}, \tag{2}$$

$$\mathbf{n} \cdot \nabla \phi = 0, \tag{3}$$

in the bulk fluid and at the solid surface. Here $\mu$ is the generalized chemical potential and $\mathbf{n}$ is the outward unit normal vector of the surface pointing from the fluid into the solid. Since the boundary conditions at the solid surface are not the focus here, we assume, for simplicity, that the solid surface has identical interaction with the two fluid components.

For an incompressible binary fluid flow, we have the incompressibility condition

$$\nabla \cdot \mathbf{v} = 0, \tag{4}$$

and the continuity equation for $\phi$,

$$\frac{\partial \phi}{\partial t} = -\nabla \cdot \mathbf{J} = -\nabla \cdot (\phi \mathbf{v} + \mathbf{j}), \tag{5}$$

in which $\mathbf{J} = \phi \mathbf{v} + \mathbf{j}$ is the total current density for the transport of $\phi$ and $\mathbf{j}$ is the diffusive current density measuring the relative motion between the two components.

For the two-phase flows of a binary fluid bounded by solid surfaces, we assume that the local equilibration at the solid surface is very fast, and hence the equilibrium boundary condition (3) for $\phi$ still applies [22,23]. In addition, there are the impermeability conditions

$$\mathbf{n} \cdot \mathbf{v} = 0, \quad \mathbf{n} \cdot \mathbf{j} = 0, \tag{6}$$

and the no-slip boundary condition for $\mathbf{v}$ at the solid surface.

Now we employ Onsager's variational principle to derive the dynamic equations for two-phase binary fluids on solid surfaces. To this end, we first find the Rayleighian defined by $\Re = \dot{F} + \Phi_F$ [6,18-24,26,27]. Here $\dot{F}$ is the rate of change of the free energy, given by

$$\dot{F} = \int \mu \frac{\partial \phi}{\partial t} d\mathbf{r} + \int K \mathbf{n} \cdot \nabla \phi \frac{\partial \phi}{\partial t} dA, \tag{7}$$

from which we obtain

$$\dot{F} = \int \nabla \mu \cdot (\phi \mathbf{v} + \mathbf{j}) d\mathbf{r}, \tag{8}$$

using the boundary condition (3), the continuity equation (5), and the impermeability conditions (6). The other part in $\Re$ is the dissipation functional $\Phi_F$, given by

$$\Phi_F = \int \frac{\eta}{4} \left[ \nabla \mathbf{v} + (\nabla \mathbf{v})^T \right]^2 d\mathbf{r} + \int \frac{\mathbf{j}^2}{2M} d\mathbf{r}, \tag{9}$$

in which the first term on the right-hand side is due to the viscous dissipation, with $\eta$ being the shear viscosity, and the second term is due to the diffusive dissipation, with $M$ being the mobility. Theoretically, both $\eta$ and $M$ may depend on the local mass fraction $\phi$, i.e. $\eta = \eta(\phi)$ and $M = M(\phi)$.

Subject to the incompressibility condition (4), minimizing the Rayleighian $\Re = \dot{F} + \Phi_F$ with respect to the rates $\mathbf{v}$ and $\mathbf{j}$ gives the momentum equation for $\mathbf{v}$ and the constitutive relation for $\mathbf{j}$, respectively:

$$\nabla \cdot \boldsymbol{\sigma} - \phi \nabla \mu = 0, \tag{10}$$

$$\mathbf{j} = -M\nabla\mu, \tag{11}$$

in which $\boldsymbol{\sigma}$ is the total stress tensor given by

$$\boldsymbol{\sigma} = -p\mathbf{I} + \boldsymbol{\sigma}_{\text{visc}}, \tag{12}$$

with $p$ being the pressure and $\boldsymbol{\sigma}_{\text{visc}}$ being the Newtonian viscous stress tensor

$$\boldsymbol{\sigma}_{\text{visc}} = \eta\left[\nabla\mathbf{v} + (\nabla\mathbf{v})^T\right]. \tag{13}$$

In summary, the dynamics of two-phase binary fluids on solid surfaces is described by the incompressibility condition (4), the continuity equation (5) for $\phi$, the momentum equation (10) for $\mathbf{v}$, supplemented with the constitutive equations (11) for $\mathbf{j}$, (12) for $\boldsymbol{\sigma}$, and (13) for $\boldsymbol{\sigma}_{\text{visc}}$. The boundary conditions applied at the solid surfaces are the local equilibrium condition (3) for $\phi$, the impermeability conditions (6) for $\mathbf{v}$ and $\mathbf{j}$, and the no-slip condition for $\mathbf{v}$.

### B. Free energy balance equation for an open system

Now we consider a particular situation in which the fluid is partially bounded by the solid surface [24]. Suppose the fluid domain under consideration is $\Omega$ and its boundary is $\partial\Omega$ which consists of solid surfaces (SS) and a few inlets and outlets (IO), i.e., $\partial\Omega = \text{SS} \cup \text{IO}$. The boundary conditions applied at the solid surface have already been presented above. We still use $\mathbf{n}$ to denote the outward pointing unit normal vector at inlets and outlets.

We start from the rate of change of the free energy (7), which takes the form of

$$\dot{F} = \int_\Omega \mu \frac{\partial\phi}{\partial t} d\mathbf{r} + \int_{\text{IO}} K\mathbf{n}\cdot\nabla\phi \frac{\partial\phi}{\partial t} dA, \tag{14}$$

in which the surface integral is only contributed by the inlets and outlets as the boundary condition (3), $\mathbf{n}\cdot\nabla\phi = 0$, is applied at the solid surface. The volume integral in $\dot{F}$ can be expressed as

$$\int_\Omega \mu \frac{\partial\phi}{\partial t} d\mathbf{r} = -\int_{\partial\Omega} \mu\mathbf{n}\cdot\mathbf{J}dA + \int_{\partial\Omega} \mathbf{n}\cdot\boldsymbol{\sigma}\cdot\mathbf{v}dA + \int_\Omega \nabla\mu\cdot\mathbf{j}d\mathbf{r} - \int_\Omega \boldsymbol{\sigma}_{\text{visc}}:\nabla\mathbf{v}d\mathbf{r}, \tag{15}$$

where we have used incompressibility condition (4), the continuity equation (5), the momentum equation (10), and integration by parts twice. There are four terms on the right-hand side and they will be explained below one by one. The first term $-\int_{\partial\Omega} \mu\mathbf{n}\cdot\mathbf{J}dA$ is the rate of the free energy pumped into the system through the inward normal current density $-\mathbf{n} \cdot \mathbf{J}$, the second term is the rate of the work done to the system by the total stress $\boldsymbol{\sigma}$ defined in Eq. (12), the third term is the rate of change of the free energy due to diffusive dissipation, and the fourth term is the rate of change of the free energy due to viscous dissipation. Note that the first two terms are surface integrals that are only contributed by the inlets and outlets as the boundary conditions $\mathbf{n} \cdot \mathbf{J} = 0$ and $\mathbf{v} = 0$ are applied at the solid surfaces. Substituting the constitutive equations (11) and (13) for $\mathbf{j}$ and $\boldsymbol{\sigma}_{\text{visc}}$ into Eq. (15), we obtain

$$\int_\Omega \mu \frac{\partial \phi}{\partial t} d\mathbf{r} = -\int_{\text{IO}} \mu \mathbf{n} \cdot \mathbf{J} dA + \int_{\text{IO}} \mathbf{n} \cdot \boldsymbol{\sigma} \cdot \mathbf{v} dA - \int_\Omega \frac{\mathbf{j}^2}{M} d\mathbf{r} - \int_\Omega \frac{\eta}{2} \left[ \nabla \mathbf{v} + (\nabla \mathbf{v})^T \right]^2 d\mathbf{r} . \quad (16)$$

For stationary states with $\partial \phi / \partial t = 0$, Eq. (16) gives the free energy balance equation

$$-\int_{\text{IO}} \mu \mathbf{n} \cdot \mathbf{J} dA + \int_{\text{IO}} \mathbf{n} \cdot \boldsymbol{\sigma} \cdot \mathbf{v} dA = \int_\Omega \frac{\mathbf{j}^2}{M} d\mathbf{r} + \int_\Omega \frac{\eta}{2} \left[ \nabla \mathbf{v} + (\nabla \mathbf{v})^T \right]^2 d\mathbf{r} . \quad (17)$$

It means that the free energy pumped into the system and the work done to the system are completely dissipated by diffusion and viscous momentum transport.

Below we consider stationary states in the proximity of the equilibrium state at which we have $\mathbf{v}_{\text{eq}} = 0$, $\mathbf{j}_{\text{eq}} = 0$, $p_{\text{eq}} = \text{const.}$, $\boldsymbol{\sigma}_{\text{eq}} = -p_{\text{eq}} \mathbf{I}$, and $\mu_{\text{eq}} = \text{const.}$, with the subscript "eq" denoting equilibrium-state properties. From $\nabla \cdot \mathbf{v} = 0$ and $\nabla \cdot \mathbf{J} = 0$ for stationary states, we obtain $\int_{\text{IO}} \mathbf{n} \cdot \mathbf{J} dA = 0$ and $\int_{\text{IO}} \mathbf{n} \cdot \mathbf{v} dA = 0$. Substituting these equilibrium-state and stationary-state properties into the free energy balance equation (17) for stationary states, we obtain

$$-\int_{\text{IO}} (\mu - \mu_{\text{eq}}) \mathbf{n} \cdot \mathbf{J} dA + \int_{\text{IO}} \mathbf{n} \cdot (\boldsymbol{\sigma} - \boldsymbol{\sigma}_{\text{eq}}) \cdot \mathbf{v} dA = \int_\Omega \frac{\mathbf{j}^2}{M} d\mathbf{r} + \int_\Omega \frac{\eta}{2} \left[ \nabla \mathbf{v} + (\nabla \mathbf{v})^T \right]^2 d\mathbf{r} , \quad (18)$$

in which $\mu - \mu_{\text{eq}}$ and $\mathbf{n} \cdot (\boldsymbol{\sigma} - \boldsymbol{\sigma}_{\text{eq}})$ are regarded as the generalized forces due to deviation from the equilibrium state. Note that they are forces acting on the system at the open boundary (inlets and outlets) and their conjugate rates (fluxes) are $\mathbf{n} \cdot \mathbf{J}$ and $\mathbf{v}$ at the boundary as well.

### C. A general formulation for the generalized Lorentz reciprocal theorem

The free energy balance equation (18) for stationary states connects the forces and their conjugate rates at the boundary (on the left-hand side) with the rates in the bulk region (on the right-hand side). This connection allows us to derive the GLRT for the boundary forces and rates from the local ORR for the forces and rates in the bulk region [8]. Note that this theorem, first derived here for a Ginzberg-Landau-type model, is of general applicability. Therefore, we present a general formulation for the GLRT with the short notations introduced as follows.

We use $F_\beta$ to denote the generalized boundary forces and $I_\beta$ to denote their conjugate generalized rates with $\beta = 1, 2, 3, \cdots$. As a result, the surface integrals in the free energy balance equation (18) become

$$-\int_{\text{IO}} \left(\mu - \mu_{\text{eq}}\right) \mathbf{n} \cdot \mathbf{J} dA + \int_{\text{IO}} \mathbf{n} \cdot \left(\boldsymbol{\sigma} - \boldsymbol{\sigma}_{\text{eq}}\right) \cdot \mathbf{v} dA = \sum_\beta F_\beta I_\beta . \qquad (19)$$

Furthermore, we use $i_m$ to denote the rates in the bulk region, with $m = 1, 2, 3, \cdots$, and the volume integrals in the free energy balance equation (18) become

$$\int_\Omega \frac{\mathbf{j}^2}{M} d\mathbf{r} + \int_\Omega \frac{\eta}{2} \left[\nabla \mathbf{v} + \left(\nabla \mathbf{v}\right)^T\right]^2 d\mathbf{r} = \sum_{m,n} i_m \varsigma_{mn} i_n . \qquad (20)$$

Here $\varsigma_{mn}$ is the resistance matrix which is symmetric (i.e., $\varsigma_{mn} = \varsigma_{nm}$) according to Onsager [6] and $\sum_{m,n} i_m \varsigma_{mn} i_n = 2\Phi_F$ is the rate of free energy dissipation in the bulk region. For convenience, we have used a discrete summation to indicate integrations at the boundary and in the bulk. Now the free energy balance equation (18) can be rewritten in a general form of

$$\sum_\beta F_\beta I_\beta = \sum_{m,n} i_m \varsigma_{mn} i_n , \text{ or equivalently } \mathbf{F}^T \mathbf{I} = \mathbf{i}^T \varsigma \mathbf{i} , \qquad (21)$$

in which $\mathbf{F}$, $\mathbf{I}$, and $\mathbf{i}$ are column vectors, the symmetric matrix $\varsigma$, with $\varsigma = \varsigma^T$, is formed by the entries $\varsigma_{mn}$, and the superscript "T" denotes the matrix transpose.

Furthermore, for the stationary states *in the proximity of the equilibrium state*, the system can be linearized with the linear relations

$$\mathbf{F} = \mathbf{A}\mathbf{i} \text{ and } \mathbf{I} = \mathbf{B}\mathbf{i} , \qquad (22)$$

in which the matrices $\mathbf{A}$ and $\mathbf{B}$ are equilibrium-state properties independent of the dynamic state. Substituting these relations into Eq. (21) and using the symmetry of $\varsigma$, we obtain

$$\varsigma = \mathbf{A}^T \mathbf{B} = \mathbf{B}^T \mathbf{A} = \varsigma^T . \qquad (23)$$

Now we consider two stationary states labelled by superscripts "(1)" and "(2)". Using Eqs. (22) and (23), we obtain the GLRT as

$$\left[\mathbf{F}^{(1)}\right]^T \mathbf{I}^{(2)} = \left[\mathbf{i}^{(1)}\right]^T \varsigma \mathbf{i}^{(2)} = \left[\mathbf{i}^{(2)}\right]^T \varsigma \mathbf{i}^{(1)} = \left[\mathbf{F}^{(2)}\right]^T \mathbf{I}^{(1)}, \tag{24}$$

in which $\mathbf{A}$, $\mathbf{B}$, and $\varsigma = \mathbf{A}^T \mathbf{B}$ are independent of the dynamic state. Equation (24) gives a general formulation of the GLRT for the boundary forces $\mathbf{F}$ and their conjugate rates $\mathbf{I}$. In particular, for the two-phase binary fluids considered here, the GLRT (24) is explicitly expressed in Eq. (A24) in the Appendix.

Finally, since the system is in the proximity of the equilibrium state and can be linearized, we have the linear kinetic equations

$$\mathbf{F} = \mathbf{R}\mathbf{I}, \tag{25}$$

in which the resistance matrix $\mathbf{R}$ is formed by the kinetic coefficients $R_{\beta\gamma}$. It follows that by using the GLRT in Eq. (24), we immediately obtain the GORR for $\mathbf{R}$,

$$\mathbf{R} = \mathbf{R}^T, \tag{26}$$

which results from the local ORR $\varsigma = \varsigma^T$ and the proximity of the equilibrium state. In the next subsection, as a specific example of the GORR (26) for the resistance matrix, we consider the GORR for the cross coupling of two transport processes in a simple capillary.

### D. Application: Cross coupling of two transport processes in a simple capillary

As a concrete example, we consider an incompressible two-phase binary fluid slowly flowing through a uniform cylindrical capillary with two open ends (inlet and outlet) [8], as shown schematically in Fig. 1. For demonstration purpose, we proceed with the following assumptions: (i) the flow in the capillary is uniaxial, i.e., the velocity $\mathbf{v}$ has only one non-zero component along the cylindrical axis, with $\mathbf{v} = (\mathbf{v} \cdot \mathbf{n})\mathbf{n}$, and (ii) the normal component of the total stress tensor $\boldsymbol{\sigma}$ in Eq. (12) is simply given by $\mathbf{n} \cdot \boldsymbol{\sigma} \cdot \mathbf{n} = -p$ at the inlet and outlet.

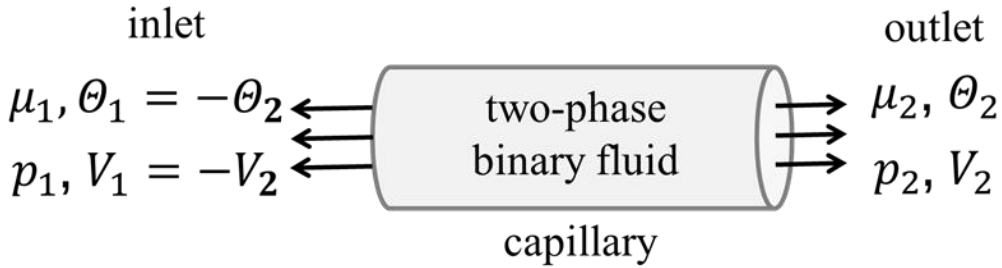

Fig. 1: Schematic illustration for the cross coupling of two transport processes in a simple capillary. The two ends of the capillary are connected to two reservoirs, respectively, where the chemical potentials ($\mu_1$ and $\mu_2$) and pressures ($p_1$ and $p_2$) are prescribed. The integrated outward flux of the solute volume is denoted by $\Theta_i$ and the integrated outward flux of the fluid volume is denoted by $V_i$.

In this flow geometry, the surface integrals in equation (19) can be reduced to

$$\sum_{\beta} F_{\beta} I_{\beta} = -\left(\mu_1 - \mu_{eq}\right)\Theta_1 - \left(\mu_2 - \mu_{eq}\right)\Theta_2 - \left(p_1 - p_{eq}\right)V_1 - \left(p_2 - p_{eq}\right)V_2, \tag{27}$$

in which $\Theta_i = \int_{IO} \mathbf{n}_i \cdot \mathbf{J}_i dA$ is the integrated outward flux of the solute volume, $V_i = \int_{IO} \mathbf{n}_i \cdot \mathbf{v}_i dA$ is the integrated outward flux of the fluid volume, and $\mathbf{n}_i$ is the outward pointing unit normal vector at the two open boundaries, with the subscripts $i = 1$ for the inlet and $i = 2$ for the outlet. Note that in stationary states, we have $\Theta_1 + \Theta_2 = 0$ and $V_1 + V_2 = 0$ from the conservation of solute volume and total volume. As a result, Eq. (27) becomes

$$\sum_{\beta} F_{\beta} I_{\beta} = \Delta\mu \Theta_2 + \Delta p V_2, \tag{28}$$

with $\Delta\mu \equiv \mu_1 - \mu_2$ and $\Delta p \equiv p_1 - p_2$. It follows that in this flow geometry, the GLRT is given by

$$\Delta\mu^{(1)}\Theta_2^{(2)} + \Delta p^{(1)}V_2^{(2)} = \Delta\mu^{(2)}\Theta_2^{(1)} + \Delta p^{(2)}V_2^{(1)}, \tag{29}$$

and the linear kinetic equation takes the form of

$$\begin{bmatrix} \Delta\mu \\ \Delta p \end{bmatrix} = \begin{bmatrix} R_{\phi\phi} & R_{\phi v} \\ R_{v\phi} & R_{vv} \end{bmatrix} \begin{bmatrix} \Theta_2 \\ V_2 \end{bmatrix}, \tag{30}$$

with the GORR given by $R_{\phi v} = R_{v\phi}$, which represents the symmetry in the cross coupling of two transport processes through the cylindrical capillary.

### III. MICROPOLAR FLUIDS

#### A. Hydrodynamic equations derived from Onsager's variational principle

Another well-studied example of complex fluids is micropolar fluids which have internal microstructure possessing its own spin. As a result, the stress tensor in continuum dynamics is no longer symmetric. Below we use Onsager's variational principle to derive the hydrodynamic equations for micropolar fluids, with inertial forces neglected in slow viscous flows [10].

The free energy functional of an isothermal micropolar fluids is given by

$$F[\rho(\mathbf{r})] = \int f(\rho) d\mathbf{r}, \tag{31}$$

in which $\rho(\mathbf{r})$ is the mass density field and $f(\rho)$ is the free energy density locally determined by $\rho$. The pressure is given by the equation of state

$$p(\rho) = \mu(\rho)\rho - f(\rho), \tag{32}$$

in which $\mu(\rho) = df(\rho)/d\rho$ is the chemical potential. For isothermal fluids, $p$ satisfies the Gibbs-Duhem relation $\nabla p = \rho \nabla \mu$ [8,18,24,25]. The continuity equation for $\rho$ is

$$\frac{\partial \rho}{\partial t} = -\nabla \cdot (\rho \mathbf{v}), \tag{33}$$

in which $\mathbf{v}$ is the fluid velocity. The hydrodynamic equations for micropolar fluids can be derived from Onsager's variational principle by minimizing the Rayleighian, $\Re = \dot{F} + \Phi_F$ [6,18-24,26,27]. Here $\dot{F}$ is the rate of change of the free energy, given by

$$\dot{F} = \int_\Omega \mu \frac{\partial \rho}{\partial t} d\mathbf{r} + \int_{\partial\Omega} \mathbf{n} \cdot \mathbf{v} f(\rho) dA, \tag{34}$$

in which $\partial\Omega$ represents the solid surface and $\mathbf{n}$ is the outward unit normal vector of the surface pointing from the fluid into the solid. For the derivation of hydrodynamic equations in the bulk region, the solid surface is assumed to be not moving, and hence $\mathbf{v} = 0$ at $\partial\Omega$ where the no-slip condition and the impermeability condition are both applied. This assumption, however, will be lifted later when moving walls are introduced. From Eq. (34), we obtain

$$\dot{F} = \int \nabla \mu \cdot (\rho \mathbf{v}) d\mathbf{r}, \tag{35}$$

with the help of the continuity equation (33) and $\mathbf{v} = 0$ at $\partial\Omega$. The other part in $\Re$ is the dissipation functional $\Phi_F$, given by

$$\Phi_F = \int \left[ \frac{1}{2}\kappa(\nabla \cdot \mathbf{v})^2 + \eta \mathbf{E}_v : \mathbf{E}_v + \frac{1}{2}\nu_1(\nabla \cdot \boldsymbol{\omega})^2 + \nu_2 \mathbf{E}_\omega : \mathbf{E}_\omega + \frac{1}{2}\xi\left(\frac{1}{2}\nabla \times \mathbf{v} - \boldsymbol{\omega}\right)^2 \right] d\mathbf{r}, \tag{36}$$

in which $\mathbf{v}$ and $\boldsymbol{\omega}$ are the velocity and spin fields, respectively. Here the symmetric, traceless rate-of-strain dyadic $\mathbf{E}_v$ and rate-of-spin-strain dyadic $\mathbf{E}_\omega$ are defined by

$$\mathbf{E}_v = \frac{1}{2}\left[\nabla \mathbf{v} + (\nabla \mathbf{v})^T\right] - \frac{1}{3}(\nabla \cdot \mathbf{v})\mathbf{I}, \tag{37}$$

$$\mathbf{E}_\omega = \frac{1}{2}\left[\nabla \boldsymbol{\omega} + (\nabla \boldsymbol{\omega})^T\right] - \frac{1}{3}(\nabla \cdot \boldsymbol{\omega})\mathbf{I}. \tag{38}$$

There are five viscosity coefficients $\kappa$, $\eta$, $v_1$, $v_2$, and $\xi$, in which $\kappa$ is the dilatational viscosity, $\eta$ is the shear viscosity, $v_1$ is the dilatational spin viscosity, $v_2$ is the shear spin viscosity, and $\xi$ is called the vortex viscosity.

Minimizing the Rayleighian $\mathfrak{R} = \dot{F} + \Phi_F$ with respect to the rates $\mathbf{v}$ and $\boldsymbol{\omega}$ gives the equations for linear and angular momentum:

$$\nabla \cdot \boldsymbol{\sigma} = 0, \tag{39}$$

$$\nabla \cdot \mathbf{C} + \boldsymbol{\Omega} = 0, \tag{40}$$

respectively, where the total stress tensor $\boldsymbol{\sigma}$ and the couple-stress tensor $\mathbf{C}$ are given by

$$\boldsymbol{\sigma} = -p\mathbf{I} + \boldsymbol{\sigma}_{visc}, \tag{41}$$

$$\mathbf{C} = v_1(\nabla \cdot \boldsymbol{\omega})\mathbf{I} + 2v_2 \mathbf{E}_\omega, \tag{42}$$

in which

$$\boldsymbol{\sigma}_{visc} = \kappa(\nabla \cdot \mathbf{v})\mathbf{I} + 2\eta \mathbf{E}_v + \frac{1}{2}\boldsymbol{\varepsilon} \cdot \boldsymbol{\Omega}, \tag{43}$$

is the viscous stress tensor, $\boldsymbol{\varepsilon}$ is the Levi-Civita symbol, and $\boldsymbol{\Omega} = \xi(\nabla \times \mathbf{v}/2 - \boldsymbol{\omega})$ is associated with the antisymmetric part of the total stress tensor $\boldsymbol{\sigma}$. Note that inertial forces are neglected in the the linear and angular momentum equations.

In summary, the dynamics of micropolar fluids on solid surfaces is described by the continuity equation (33) for $\rho$, the linear momentum equation (39) for $\mathbf{v}$, and the angular momentum equation (40) for $\boldsymbol{\omega}$, supplemented with the equation of state (32) for $p$ and the constitutive equations (41) for $\boldsymbol{\sigma}$, (42) for $\mathbf{C}$, and (43) for $\boldsymbol{\sigma}_{visc}$. The boundary conditions applied at the solid surface are the impermeability condition for $\mathbf{v}$ and the no-slip conditions for $\mathbf{v}$ and $\boldsymbol{\omega}$.

## B. Free energy balance equation for a system with moving solid boundary

Now we consider the situation in which the solid surface is *moving*. Suppose the volume domain of the micropolar fluid under consideration is $\Omega$ and its boundary is $\partial\Omega$. The rate of change of the free energy is given by Eq. (34) with $\mathbf{n}\cdot\mathbf{v} \neq 0$ at the solid surface that is moving. Using the continuity equation (33) and the equation of state (32), we have

$$\dot{F} = -\int_{\partial\Omega} \mathbf{n}\cdot\mathbf{v} p(\rho) dA + \int_{\Omega} \nabla p \cdot \mathbf{v} d\mathbf{r}, \qquad (44)$$

which can be rewritten as

$$\dot{F} = \int_{\partial\Omega} \mathbf{n}\cdot\boldsymbol{\sigma}\cdot\mathbf{v} dA - \int_{\Omega} \boldsymbol{\sigma}_{visc} : \nabla \mathbf{v} d\mathbf{r}, \qquad (45)$$

with the help of Eqs. (39) and (41). Furthermore, using Eq. (43), we can rewrite the volume integral in Eq. (45) as

$$\int_{\Omega} \boldsymbol{\sigma}_{visc} : \nabla \mathbf{v} d\mathbf{r} =$$
$$-\int_{\partial\Omega} \mathbf{n}\cdot\mathbf{C}\cdot\boldsymbol{\omega} dA + \int_{\Omega} \left[ \kappa(\nabla\cdot\mathbf{v})^2 + 2\eta \mathbf{E}_v : \mathbf{E}_v + \nu_1(\nabla\cdot\boldsymbol{\omega})^2 + 2\nu_2 \mathbf{E}_\omega : \mathbf{E}_\omega + \xi\left(\frac{1}{2}\nabla\times\mathbf{v} - \boldsymbol{\omega}\right)^2 \right] d\mathbf{r}. \qquad (46)$$

It follows that $\dot{F}$ in Eq. (45) can be expressed in a more illustrative form

$$\dot{F} = \int_{\partial\Omega} \mathbf{n}\cdot\boldsymbol{\sigma}\cdot\mathbf{v} dA + \int_{\partial\Omega} \mathbf{n}\cdot\mathbf{C}\cdot\boldsymbol{\omega} dA$$
$$- \int_{\Omega} \left[ \kappa(\nabla\cdot\mathbf{v})^2 + 2\eta \mathbf{E}_v : \mathbf{E}_v + \nu_1(\nabla\cdot\boldsymbol{\omega})^2 + 2\nu_2 \mathbf{E}_\omega : \mathbf{E}_\omega + \xi\left(\frac{1}{2}\nabla\times\mathbf{v} - \boldsymbol{\omega}\right)^2 \right] d\mathbf{r}, \qquad (47)$$

in which the two surface integrals represent the work done to the fluid at $\partial\Omega$ and the volume integral equals $2\Phi_F$ in $\Omega$, representing the rate of viscous dissipation in the bulk region.

Below we consider stationary states in the proximity of the equilibrium state at which we have $\mathbf{v}_{eq} = 0$, $\boldsymbol{\omega}_{eq} = 0$, $p_{eq} = \text{const.}$, $\boldsymbol{\sigma}_{eq} = -p_{eq}\mathbf{I}$, and $\mathbf{C}_{eq} = 0$, with the subscript "eq" denoting equilibrium-state properties. For stationary states with $\partial\rho/\partial t = 0$, if they are close to the equilibrium state with $\rho_{eq} = \text{const.}$ (due to $p_{eq} = \text{const.}$), then we have $\nabla\cdot\mathbf{v} = 0$ from the continuity equation (33). Combining $\nabla\cdot\mathbf{v} = 0$ and $\dot{F} = -\int_{\Omega} p\nabla\cdot\mathbf{v} d\mathbf{r}$ from Eq. (44), we obtain $\dot{F} = 0$ for these stationary states. (It is interesting to note that due to the moving boundary, we have $\dot{F} = 0$ only for these near-equilibrium stationary states with $\nabla\cdot\mathbf{v} \approx 0$.) As a result, we obtain

$$\int_{\partial\Omega} \mathbf{n}\cdot\boldsymbol{\sigma}\cdot\mathbf{v}\,dA + \int_{\partial\Omega} \mathbf{n}\cdot\mathbf{C}\cdot\boldsymbol{\omega}\,dA = \int_{\Omega}\left[2\eta\mathbf{E}_v{:}\mathbf{E}_v + \nu_1(\nabla\cdot\boldsymbol{\omega})^2 + 2\nu_2\mathbf{E}_\omega{:}\mathbf{E}_\omega + \xi\left(\frac{1}{2}\nabla\times\mathbf{v} - \boldsymbol{\omega}\right)^2\right]d\mathbf{r}, \quad (48)$$

which is the free energy balance equation for stationary states in the proximity of the equilibrium state. Using $p_{eq} = \text{const.}$, $\boldsymbol{\sigma}_{eq} = -p_{eq}\mathbf{I}$, $\mathbf{C}_{eq} = 0$ and $\nabla\cdot\mathbf{v} \approx 0$ and thus $\int_{\partial\Omega}\mathbf{n}\cdot\mathbf{v}\,dA = 0$, we have

$$\int_{\partial\Omega} \mathbf{n}\cdot(\boldsymbol{\sigma}-\boldsymbol{\sigma}_{eq})\cdot\mathbf{v}\,dA + \int_{\partial\Omega} \mathbf{n}\cdot(\mathbf{C}-\mathbf{C}_{eq})\cdot\boldsymbol{\omega}\,dA =$$
$$\int_{\Omega}\left[2\eta\mathbf{E}_v{:}\mathbf{E}_v + \nu_1(\nabla\cdot\boldsymbol{\omega})^2 + 2\nu_2\mathbf{E}_\omega{:}\mathbf{E}_\omega + \xi\left(\frac{1}{2}\nabla\times\mathbf{v} - \boldsymbol{\omega}\right)^2\right]d\mathbf{r}, \quad (49)$$

in which $\mathbf{n}\cdot(\boldsymbol{\sigma}-\boldsymbol{\sigma}_{eq})$ and $\mathbf{n}\cdot(\mathbf{C}-\mathbf{C}_{eq})$ are regarded as the generalized forces due to deviation from the equilibrium state. Note that they are forces acting on the system at the moving solid boundary and their conjugate rates (fluxes) are $\mathbf{v}$ and $\boldsymbol{\omega}$ at the boundary as well.

According to the discussion in Sec. II.C, the stationary-state free energy balance equation (49) can be cast into the general form of Eq. (21) as $\sum_\beta F_\beta I_\beta = \sum_{m,n} i_m \varsigma_{mn} i_n$, with

$$\sum_\beta F_\beta I_\beta \equiv \int_{\partial\Omega} \mathbf{n}\cdot(\boldsymbol{\sigma}-\boldsymbol{\sigma}_{eq})\cdot\mathbf{v}\,dA + \int_{\partial\Omega} \mathbf{n}\cdot(\mathbf{C}-\mathbf{C}_{eq})\cdot\boldsymbol{\omega}\,dA \quad (50)$$

at the moving solid boundary and

$$\sum_{m,n} i_m \varsigma_{mn} i_n \equiv \int_{\Omega}\left[2\eta\mathbf{E}_v{:}\mathbf{E}_v + \nu_1(\nabla\cdot\boldsymbol{\omega})^2 + 2\nu_2\mathbf{E}_\omega{:}\mathbf{E}_\omega + \xi\left(\frac{1}{2}\nabla\times\mathbf{v} - \boldsymbol{\omega}\right)^2\right]d\mathbf{r} \quad (51)$$

in the bulk region. Here we are ready to make use of the discussion and formulation presented in Sec. IIC for micropolar fluids with moving solid boundary. In particular, with the boundary forces and their conjugate rates identified in Eq. (50), we obtain the corresponding GLRT (24) and GORR (26) for the resistance matrix in the linear response relation (25) for generalized boundary forces and fluxes. The GLRT (24) is explicitly expressed in Eq. (A27) in the Appendix. In the next subsection, as a specific example of the GORR (26) for the resistance matrix, we consider the GORR for the translation and rotation of a rigid body in a quiescent micropolar fluid.

### C. Application: Translation and rotation of a rigid body in a quiescent micropolar fluid

As an application of the GLRT and GORR, we now consider the slow viscous flow induced by the translation and rotation of a rigid body in a quiescent micropolar fluid, as shown schematically in Fig. 2. This example has been discussed in Ref. [10]. Using our general notations, we briefly reproduce their results below to make our presentation more complete and self-contained.

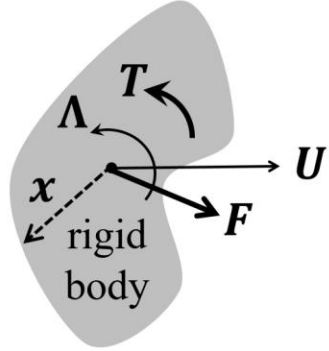

Fig. 2: Schematic illustration for the translation and rotation of a rigid body with an arbitrary shape in a quiescent micropolar fluid. Here $\mathbf{U}$ and $\mathbf{\Lambda}$ are the translational center-of-mass velocity and angular velocity of the rigid body, respectively, and $\mathbf{F}$ and $\mathbf{T}$ are the hydrodynamic force and torque (w.r.t. center-of-mass) exerted by the rigid body on the fluid, respectively.

We consider a rigid body that is moving with the translational center-of-mass velocity $\mathbf{U}$ and rotating with the angular velocity $\mathbf{\Lambda}$. The fluid velocity at the point $\mathbf{x}$ (measured relative to the center-of-mass) at the surface of the body is $\mathbf{v} = \mathbf{U} + \mathbf{\Lambda} \times \mathbf{x}$ according to the no-slip boundary condition. As a result, the surface integrals in Eq. (50) become

$$\sum_\beta F_\beta I_\beta = \mathbf{F} \cdot \mathbf{U} + \mathbf{T} \cdot \mathbf{\Lambda}, \tag{52}$$

with $\mathbf{F} = \int_{\partial\Omega} \mathbf{n} \cdot (\boldsymbol{\sigma} - \boldsymbol{\sigma}_{eq}) dA$ and $\mathbf{T} = \int_{\partial\Omega} \{\mathbf{x} \times [\mathbf{n} \cdot (\boldsymbol{\sigma} - \boldsymbol{\sigma}_{eq})] + \mathbf{n} \cdot (\mathbf{C} - \mathbf{C}_{eq})\} dA$ being the hydrodynamic force and torque (with respect to the center-of-mass) exerted by the rigid body on the fluid. Here the no-slip condition for the spin field, $\boldsymbol{\omega} = \mathbf{\Lambda}$, is used for $\boldsymbol{\omega}$ at $\partial\Omega$ in Eq. (50). It follows that in the present application, the GLRT is given by

$$\mathbf{F}^{(1)} \cdot \mathbf{U}^{(2)} + \mathbf{T}^{(1)} \cdot \mathbf{\Lambda}^{(2)} = \mathbf{F}^{(2)} \cdot \mathbf{U}^{(1)} + \mathbf{T}^{(2)} \cdot \mathbf{\Lambda}^{(1)}, \tag{53}$$

and the linear kinetic equation takes the form of

$$\begin{bmatrix} \mathbf{F} \\ \mathbf{T} \end{bmatrix} = \begin{bmatrix} \mathbf{R}_{TT} & \mathbf{R}_{TR} \\ \mathbf{R}_{RT} & \mathbf{R}_{RR} \end{bmatrix} \cdot \begin{bmatrix} \mathbf{U} \\ \mathbf{\Lambda} \end{bmatrix}, \tag{54}$$

with

$$\mathbf{R}_{TT} = \mathbf{R}_{TT}^T, \quad \mathbf{R}_{RR} = \mathbf{R}_{RR}^T, \quad \mathbf{R}_{TR} = \mathbf{R}_{RT}^T, \tag{55}$$

which represent the reciprocal symmetry in the cross coupling of translation and rotation of the rigid body in micropolar fluid.

## IV. THERMAL CONDUCTION IN SOLIDS AND STILL FLUIDS

In the two previous sections, we have derived the GLRT for two fluid systems under the following conditions. (i) The hydrodynamic equations are derived from Onsager's variational principle with the reciprocal symmetry reflected in local constitutive equations; and (ii) the model system is linearized in the immediate proximity of equilibrium state and hence the friction coefficients are taken as equilibrium-state properties. We would like to point out that although the systems treated above are isothermal, our approach can be readily generalized to non-isothermal systems provided that the reciprocal symmetry is preserved in local constitutive equations and the proximity of equilibrium state is ensured [8,23,24]. In this section, we consider, as an example, the thermal conduction in a solid that can be inhomogeneous and anisotropic. The discussion also applies to the thermal conduction in a still fluid.

### A. Thermal conduction equation and entropy balance equation

Consider an inhomogeneous and anisotropic solid that occupies a *fixed* domain in space denoted by $\Omega$. The boundary of $\Omega$ is denoted by $\partial\Omega$ at which heat transfer occurs between the system and its surrounding environment. According to the conservation of energy, we have

$$\dot{E} = -\int_{\partial\Omega} \mathbf{n} \cdot \mathbf{q} dA, \quad \text{or} \quad \frac{\partial e}{\partial t} = -\nabla \cdot \mathbf{q}, \tag{56}$$

in which $E = \int_{\Omega} e d\mathbf{r}$ is the internal energy of the solid, $e$ is the internal energy density, $\mathbf{q}$ is the heat current density, and $\mathbf{n}$ is the outward unit normal vector of the surface pointing from the solid into its surrounding environment. For simplicity, the thermal expansion of the solid is assumed to be negligible, and the energy equation (56) becomes

$$c_v \frac{\partial T}{\partial t} = -\nabla \cdot \mathbf{q}, \tag{57}$$

in which $c_v$ is the specific heat capacity at constant volume. The total entropy of the system is given by $S = \int_\Omega s(e) d\mathbf{r}$, in which the entropy density $s = s(e)$ is a function of the local energy density $e$. Using $ds/de = 1/T$, Eq. (56) and an integration by parts, we obtain the rate of change of the total entropy

$$\dot{S} = \int_\Omega \frac{ds}{de} \frac{\partial e}{\partial t} d\mathbf{r} = -\int_{\partial\Omega} \frac{1}{T} \mathbf{n} \cdot \mathbf{q} dA + \int_\Omega \mathbf{q} \cdot \nabla \frac{1}{T} d\mathbf{r}, \tag{58}$$

In order to apply Onsager's variational principle to non-isothermal systems (see Eq. (A12) in the Appendix), we need to find the Onsager-Machlup functional [6,8,24], $\mathscr{O} = \dot{S} + \dot{S}^* - \Phi_S$, in which $\dot{S}^*$ is the outgoing entropy flux (from the system to its surrounding environment) and $\Phi_S$ is the dissipation functional which is half the rate of entropy production. In the present case, we have $\dot{S}^* = \int_{\partial\Omega} \frac{1}{T} \mathbf{n} \cdot \mathbf{q} dA$ and $\Phi_S = \frac{1}{2} \int_\Omega \mathbf{q} \cdot \boldsymbol{\lambda}^{-1} \cdot \mathbf{q} \, d\mathbf{r}$, and hence by using Eq. (58) for $\dot{S}$, we obtain

$$\mathscr{O}[\mathbf{q}] = \int_\Omega \mathbf{q} \cdot \nabla \frac{1}{T} d\mathbf{r} - \frac{1}{2} \int_\Omega \mathbf{q} \cdot \boldsymbol{\lambda}^{-1} \cdot \mathbf{q} \, d\mathbf{r}. \tag{59}$$

Here $\boldsymbol{\lambda}$ is a symmetric and positive definite tensor, a local property that may change with the local temperature, i.e. $\boldsymbol{\lambda} = \boldsymbol{\lambda}(T)$. In addition, $\boldsymbol{\lambda}$ can vary in space due to the inhomogeneity of the solid. Therefore, we have $\boldsymbol{\lambda} = \boldsymbol{\lambda}(T(\mathbf{r}), \mathbf{r})$ in general.

Minimizing the Onsager-Machlup functional $\mathscr{O}$ in Eq. (59) with respect to $\mathbf{q}$, we obtain the constitutive relation

$$\mathbf{q} = \boldsymbol{\lambda} \cdot \nabla \frac{1}{T}, \tag{60}$$

in which $\boldsymbol{\lambda}$ can be directly related to the thermal conductivity tensor. Note that substituting Eq. (60) into $\dot{S} + \dot{S}^*$, we obtain the entropy balance equation $\dot{S} + \dot{S}^* = 2\Phi_S$. In particular, for solids at stationary states (with $\dot{S} = 0$) that are close to the equilibrium state with a homogeneous temperature $T_{eq} = $ const., the entropy balance equation becomes $\dot{S}^* = 2\Phi_S$, which can be written as

$$\int_{\partial\Omega} \left( \frac{1}{T} - \frac{1}{T_{eq}} \right) \mathbf{n} \cdot \mathbf{q} dA = \int_\Omega \mathbf{q} \cdot \boldsymbol{\lambda}^{-1} \cdot \mathbf{q} \, d\mathbf{r}, \tag{61}$$

in which $\int_{\partial\Omega} \mathbf{n}\cdot\mathbf{q}dA = 0$ is used for stationary states and $\lambda = \lambda(T_{eq},\mathbf{r})$ may still vary in space due to the inhomogeneity of the solid. Here $1/T - 1/T_{eq}$ is a thermodynamic force due to deviation from the equilibrium state. Note this force is acting on the system at the boundary $\partial\Omega$ and its conjugate rate (flux) is $\mathbf{n}\cdot\mathbf{q}$ at the boundary as well.

According to the discussion in Sec. II.C, the stationary-state entropy balance equation (61) can be cast in the general form of Eq. (21) as $\sum_\beta F_\beta I_\beta = \sum_{m,n} i_m \varsigma_{mn} i_n$, with

$$\sum_\beta F_\beta I_\beta \equiv \int_{\partial\Omega} \left( \frac{1}{T} - \frac{1}{T_{eq}} \right) \mathbf{n}\cdot\mathbf{q}dA, \tag{62}$$

at the boundary and

$$\sum_{m,n} i_m \varsigma_{mn} i_n \equiv \int_\Omega \mathbf{q}\cdot\boldsymbol{\lambda}^{-1}\cdot\mathbf{q}\, d\mathbf{r}. \tag{63}$$

in the bulk region. Here we are ready to make use of the discussion and formulation presented in Sec. IIC for thermal conduction in inhomogeneous and anisotropic solids. In particular, with the boundary forces and their conjugate rates identified in Eq. (62), we obtain the corresponding GLRT (24) and GORR (26) for the resistance matrix in the linear response relation (25) for generalized boundary forces and fluxes. The GLRT (24) is explicitly expressed in Eq. (A30) in the Appendix. In the next subsection, as a specific example of the GORR (26) for the resistance matrix, we consider the GORR for the heat transport in a thermally isolated solid with three open ports

### B. Application: Heat transport in a thermally isolated solid with three open ports

As an application of the above theoretical results, we consider the heat transport in a solid that is thermally isolated from its surrounding environment except at the three open ports, as shown in Fig. 3.

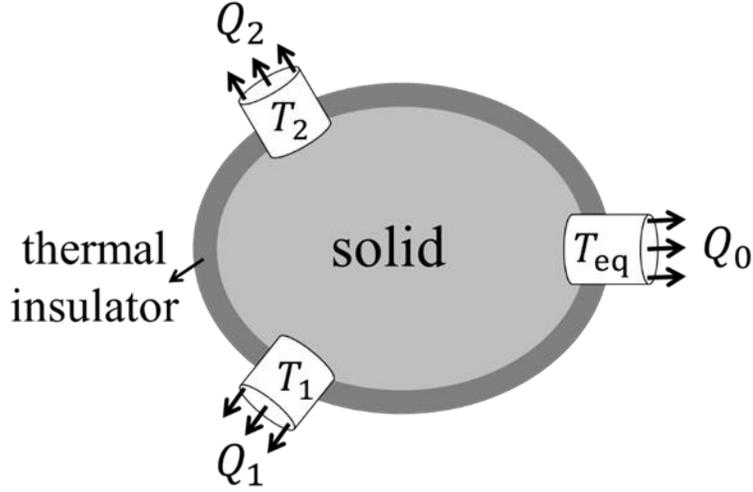

Fig. 3: Schematic illustration for the heat conduction in a solid that is thermally isolated from its surrounding environment except at the three open ports. Each open port is connected to a reservoir where the temperature is fixed, with $T = T_{eq}$ at the zeroth port, $T_1$ at the first port, and $T_2$ at the second port. The integrated outward heat flux is denoted by $Q_i$ at the $i$-th open port.

In this geometry, we have $Q_0 + Q_1 + Q_2 = 0$ for stationary states, with the subscript $i = 0, 1, 2$ labeling the three open ports and $Q_i = \int \mathbf{n}_i \cdot \mathbf{q}_i dA$ being the integrated outward heat flux at the $i$-th open port. It is obvious that there are only two independent heat fluxes for stationary states of heat transport. As a result, the surface integral in Eq. (62) becomes

$$\sum_\beta F_\beta I_\beta = \left(\frac{1}{T_1} - \frac{1}{T_{eq}}\right) Q_1 + \left(\frac{1}{T_2} - \frac{1}{T_{eq}}\right) Q_2. \tag{64}$$

It follows that in the present geometry, the GLRT is given by

$$\left(\frac{1}{T_1^{(1)}} - \frac{1}{T_{eq}}\right) Q_1^{(2)} + \left(\frac{1}{T_2^{(1)}} - \frac{1}{T_{eq}}\right) Q_2^{(2)} = \left(\frac{1}{T_1^{(2)}} - \frac{1}{T_{eq}}\right) Q_1^{(1)} + \left(\frac{1}{T_2^{(2)}} - \frac{1}{T_{eq}}\right) Q_2^{(1)}, \tag{65}$$

and the linear constitutive equation takes the form of

$$\begin{bmatrix} 1/T_1 - 1/T_{eq} \\ 1/T_2 - 1/T_{eq} \end{bmatrix} = \begin{bmatrix} R_{11} & R_{12} \\ R_{21} & R_{22} \end{bmatrix} \begin{bmatrix} Q_1 \\ Q_2 \end{bmatrix}, \tag{66}$$

with the GORR given by $R_{12} = R_{21}$, which represents the symmetry in the cross coupling of the two heat transport processes through open ports 1 and 2.

## V. CONCLUDING REMARKS

In summary, we have derived the GLRT and GORR for slow variables at the boundary of a system. We have presented three particular systems, namely, a two-phase binary fluid with open boundary, a micropolar fluid with moving solid boundary, and thermal conduction in a solid with thermally conductive boundary. For each case, we first derive the local dynamic equations using Onsager's variational principle, with Onsager's reciprocal symmetry naturally preserved in the local constitutive equations. We then show that there are two conditions that are essential to deriving the GLRT and GORR. (i) The system should be at quasi-stationary states controlled by the boundary variables whose evolution is much slower than the relaxation of the system. (ii) These quasi-stationary states should be in the close proximity of equilibrium state such that the local dynamic equations can be linearized with the phenomenological coefficients taken as equilibrium properties.

Finally, we make some remarks on the two conditions to derive the GLRT and GORR as follows:

(i) The quasi-stationary states require a separation of time scales in the system: the time evolution of the boundary variables is much slower than the relaxation in the system [18].

(ii) The close-to-equilibrium condition dictates that in the local dynamic equations, the phenomenological coefficients are treated as equilibrium properties that are independent of the dynamic state.

(iii) Under the above two conditions, our approach is independent of whether the system is isothermal or non-isothermal. Work on generalized applications to non-isothermal fluids is currently underway [8,23,24].

(iv) The above two conditions can provide some practical guidance for future experiments involving more complicated structured fluids including active matter. They will be useful for choosing slow variables and formulating ORR properly in the coarse-grained modeling of complex fluids. In particular, if the above two conditions are met, then the GLRT and GORR can be employed as a criterion for the well-posedness of the system of local dynamic equations, e.g. the equation system describing the electro-osmosis in electrolyte [27,28].

## ACKNOWLEDGEMENTS

T.Q. was supported by Hong Kong RGC CRF grant No. C1018-17G. X.X. was supported by Guangdong Technion–Israel Institute of Technology.

# APPENDIX A: ONSAGER'S RECIPROCAL RELATIONS

We consider a *closed* system described by a set of (macroscopic) state variables $\{\alpha_i\}$ with $i = 1,...,n$, measured relative to their most probable (equilibrium) values [6, 23, 24]. The entropy of the system $S$ has a maximum $S_e$ at equilibrium and $\Delta S = S - S_e$ can be expressed in the quadratic form of

$$\Delta S(\alpha_1,...,\alpha_n) = -\frac{1}{2}\sum_{i,j=1}^{n} \beta_{ij}\alpha_i\alpha_j, \tag{A1}$$

in which $\beta$ is symmetric and positive definite. The probability density at $\{\alpha_i\}$ is given by

$$f(\alpha_1,...,\alpha_n) = f(0,...,0)e^{\Delta S/k_B}, \tag{A2}$$

in which $k_B$ is the Boltzmann constant. The thermodynamic force conjugate to $\alpha_i$ is defined by

$$X_i = \frac{\partial \Delta S}{\partial \alpha_i} = -\sum_{j=1}^{n}\beta_{ij}\alpha_j, \tag{A3}$$

which is a linear combination of $\{\alpha_i\}$ not far from equilibrium.

Following the above definition of the thermodynamic forces, the equilibrium average of $\alpha_i X_j$ over the distribution function $f(\alpha_1,...,\alpha_n)$ is given by

$$\langle \alpha_i X_j \rangle = -k_B \delta_{ij}. \tag{A4}$$

The microscopic reversibility leads to the equality

$$\langle \alpha_i(t)\alpha_j(t+\tau) \rangle = \langle \alpha_j(t)\alpha_i(t+\tau) \rangle \tag{A5}$$

for time correlation functions. In the proximity of equilibrium, the macroscopic variables $\{\alpha_i\}$ satisfy the linear kinetic equations

$$\dot{\alpha}_i(t) = \sum_{j=1}^{n} L_{ij} X_j(t), \tag{A6}$$

in which $L_{ij}$ are the kinetic coefficients which form a positive definite matrix following the second law of thermodynamics. According to Onsager, fluctuations of the state variables $\{\alpha_i\}$ evolve in the mean following the same kinetic equations. Therefore, for the correlation function $\langle \alpha_i(t)\alpha_j(t+\tau)\rangle$ with $\tau$ being a time interval that is macroscopically short but microscopically long, $\alpha_j(t+\tau)$ is given by

$$\alpha_j(t+\tau) = \alpha_j(t) + \tau \dot{\alpha}_j(t) = \alpha_j(t) + \tau \sum_{k=1}^{n} L_{jk} X_k(t). \tag{A7}$$

It is worth pointing out that $\tau$ is macroscopically short for the linear expansion but microscopically long for the applicability of the kinetic equations. It follows that $\langle \alpha_i(t)\alpha_j(t+\tau)\rangle$ is given by

$$\langle \alpha_i(t)\alpha_j(t+\tau)\rangle = \langle \alpha_i(t)\alpha_j(t)\rangle + \tau \sum_{k=1}^{n} L_{jk} \langle \alpha_i(t) X_k(t)\rangle = \langle \alpha_i(t)\alpha_j(t)\rangle - \tau k_B L_{ji}. \tag{A8}$$

Similarly, $\langle \alpha_j(t)\alpha_i(t+\tau)\rangle$ is given by

$$\langle \alpha_j(t)\alpha_i(t+\tau)\rangle = \langle \alpha_j(t)\alpha_i(t)\rangle - \tau k_B L_{ij}. \tag{A9}$$

Making the above two correlation functions equal and using $\langle \alpha_i(t)\alpha_j(t)\rangle = \langle \alpha_j(t)\alpha_i(t)\rangle$ by definition, we obtain

$$L_{ji} = L_{ij} \tag{A10}$$

for the reciprocal symmetry of kinetic coefficients.

## APPENDIX B: ONSAGER'S VARIATIONAL PRINCIPLE

Based on Onsager's reciprocal symmetry for the kinetic coefficients, a variational principle can be formulated to derive the linear kinetic equations [6, 23, 24]. For this purpose, we introduce $\dot{S}(\alpha,\dot{\alpha}) = \sum_{i=1}^{n} X_i \dot{\alpha}_i$ as the rate of change of entropy and $\Phi_S(\dot{\alpha},\dot{\alpha}) = \frac{1}{2}\sum_{i,j=1}^{n} R_{ij} \dot{\alpha}_i \dot{\alpha}_j$ as the dissipation function which is half the rate of entropy production. Here $\dot{S}(\alpha,\dot{\alpha})$ is linear in the rates $\{\dot{\alpha}_i\}$, $\Phi_S(\dot{\alpha},\dot{\alpha})$ is quadratic in the rates, and the friction coefficients $R_{ij}$ are determined from the kinetic coefficients $L_{ij}$ through the relation $\sum_{j=1}^{n} L_{ij} R_{jk} = \delta_{ik}$. Maximizing the action function

$$\dot{S}(\alpha,\dot{\alpha}) - \Phi_S(\dot{\alpha},\dot{\alpha}) = \sum_{i=1}^{n} X_i \dot{\alpha}_i - \frac{1}{2} \sum_{i,j=1}^{n} R_{ij} \dot{\alpha}_i \dot{\alpha}_j \tag{A11}$$

with respect to the rates $\{\dot{\alpha}_i\}$, we obtain the kinetic equations for the time evolution of $\{\alpha_i\}$. Note that we have $R_{ji} = R_{ij}$ for the reciprocal symmetry of friction coefficients. For an open system, the action function becomes

$$\mathcal{O} = \dot{S}(\alpha,\dot{\alpha}) + \dot{S}^*(\alpha,\dot{\alpha}) - \Phi_S(\dot{\alpha},\dot{\alpha}), \tag{A12}$$

referred to as the Onsager-Machlup function, in which $\dot{S}^*$ is the rate of entropy given off by the system to the environment.

If the system is isothermal and in thermal equilibrium with the environment, then we have $\dot{S}^*(\alpha,\dot{\alpha}) = -\dot{Q}/T = -\dot{U}/T$, in which $T$ is the temperature, $\dot{Q}$ is the rate of heat transfer from the environment to the system, and $\dot{U}$ is the rate of change of the internal energy of the system, with $\dot{U} = \dot{Q}$ according to the first law of thermodynamics. A new action can be introduced as

$$-T\left[\dot{S}(\alpha,\dot{\alpha}) + \dot{S}^*(\alpha,\dot{\alpha}) - \Phi_S(\dot{\alpha},\dot{\alpha})\right] = \dot{F}(\alpha,\dot{\alpha}) + \Phi_F(\dot{\alpha},\dot{\alpha}), \tag{A13}$$

which is sometime referred to as the Rayleighian, in which $\dot{F} = \dot{U} - T\dot{S}$ is the rate of change of the Helmholtz free energy of the system, and $\Phi_F(\dot{\alpha},\dot{\alpha}) = T\Phi_S(\dot{\alpha},\dot{\alpha})$ is half the rate of free-energy dissipation. The Rayleighian can be expressed as

$$\sum_{i=1}^{n} \frac{\partial F}{\partial \alpha_i} \dot{\alpha}_i + \frac{1}{2} \sum_{i,j=1}^{n} \varsigma_{ij} \dot{\alpha}_i \dot{\alpha}_j, \tag{A14}$$

in which the first term is $\dot{F}$ and the second term is $\Phi_F$ with $\varsigma_{ji} = \varsigma_{ij}$ for the reciprocal symmetry of the friction coefficients $\varsigma_{ij}$. Minimizing the Rayleighian with respect to the rates $\{\dot{\alpha}_i\}$, we obtain the kinetic equations

$$-\frac{\partial F}{\partial \alpha_i} = \sum_{j=1}^{n} \varsigma_{ij} \dot{\alpha}_j \tag{A15}$$

for the time evolution of $\{\alpha_i\}$. Physically, these equations describe the balance between the reversible force $-\partial F / \partial \alpha_i$ and the dissipative force linear in $\{\dot{\alpha}_j\}$.

## APPENDIX C: THE LORENTZ RECIPROCAL THEOREM

### 1. Newtonian fluids

We start from the classical LRT for slow viscous flows of incompressible Newtonian fluids under isothermal condition [1-4]. Consider a Newtonian fluid in a volume region $\Omega$ with a solid boundary denoted by $\partial\Omega$. Suppose there are two flow fields $\mathbf{v}^{(1)}$ and $\mathbf{v}^{(2)}$ which are solutions of the Stokes equation for slow viscous flows, subject to the impermeability condition and the no-slip condition at the solid surface. The corresponding total stress fields are $\boldsymbol{\sigma}^{(1)}$ and $\boldsymbol{\sigma}^{(2)}$, respectively. Using the constitutive equation for the total stress tensor $\boldsymbol{\sigma} = -p\mathbf{I} + \eta\left[\nabla\mathbf{v} + (\nabla\mathbf{v})^T\right]$, and the incompressibility condition $\nabla\cdot\mathbf{v}=0$, we can obtain

$$\int_{\partial\Omega} \mathbf{n}\cdot\boldsymbol{\sigma}^{(1)}\cdot\mathbf{v}^{(2)} dA = \int_\Omega \frac{\eta}{2}\left[\nabla\mathbf{v}^{(1)} + \left(\nabla\mathbf{v}^{(1)}\right)^T\right] : \left[\nabla\mathbf{v}^{(2)} + \left(\nabla\mathbf{v}^{(2)}\right)^T\right] d\mathbf{r}. \tag{A16}$$

in which we have employed the divergence theorem and the Stokes equation $\nabla\cdot\boldsymbol{\sigma}=0$. Since the shear viscosity $\eta$ is a constant independent of the flow, Eq. (A16) is permutable for the two flow fields. From this permutation symmetry, we can obtain the LRT

$$\int_{\partial\Omega} \mathbf{n}\cdot\boldsymbol{\sigma}^{(1)}\cdot\mathbf{v}^{(2)} dA = \int_{\partial\Omega} \mathbf{n}\cdot\boldsymbol{\sigma}^{(2)}\cdot\mathbf{v}^{(1)} dA, \tag{A17}$$

in which $\mathbf{n}$ is the outward unit normal vector of the surface pointing from the fluid into the solid. Furthermore, if the no-slip condition is applied at the solid surface $\partial\Omega$, then the fluid velocity $\mathbf{v}$ is equal to the solid velocity $\mathbf{w}$, and Eq. (A17) becomes

$$\int_{\partial\Omega} \mathbf{n}\cdot\boldsymbol{\sigma}^{(1)}\cdot\mathbf{w}^{(2)} dA = \int_{\partial\Omega} \mathbf{n}\cdot\boldsymbol{\sigma}^{(2)}\cdot\mathbf{w}^{(1)} dA. \tag{A18}$$

In our previous work [15], we have shown that Eq. (A18) still holds if the no-slip condition at the solid surface is replaced by the Navier slip condition.

For a solid particle moving in the fluid, the integral $\int_{\partial\Omega}\mathbf{n}\cdot\boldsymbol{\sigma}\cdot\mathbf{w}\, dA$ at the particle surface can be written as $\sum_k F_k \dot{x}_k$, in which $\dot{x}_k$ are the generalized velocities of the solid particle and $F_k$ are the generalized forces (exerted by the particle on the fluid) conjugate to $\dot{x}_k$. The LRT (A18) then takes the form of

$$\sum_k F_k^{(1)} \dot{x}_k^{(2)} = \sum_k F_k^{(2)} \dot{x}_k^{(1)}, \tag{A19}$$

which is in the same form as Eq. (24) for the GLRT in Sec. IIC. The linearity of the Stokes equation leads to the linear response which can be expressed as

$$F_k = \sum_l \varsigma_{kl} \dot{x}_l, \tag{A20}$$

in which $\varsigma_{kl}$ are the friction coefficients which form a positive definite matrix. The reciprocal theorem in Eq. (A19) immediately gives rise to the reciprocal symmetry

$$\varsigma_{kl} = \varsigma_{lk}. \tag{A21}$$

### 2. Two-phase binary fluids

We now turn to two-phase binary fluids, which constitute a typical example of complex fluids with two-phase interfacial structures [22, 23]. We assume that the fluids are incompressible and the flows are slow (with negligible inertia) and isothermal. Consider a two-phase binary fluid in a volume region $\Omega$ with a boundary denoted by $\partial \Omega$ which consists of solid surfaces (SS) and a few inlets and outlets (IO), i.e. $\partial \Omega = \text{SS} \cup \text{IO}$. Suppose there are two stationary state solutions $\left( \mathbf{v}^{(1)}, \phi^{(1)} \right)$ and $\left( \mathbf{v}^{(2)}, \phi^{(2)} \right)$ for the dynamic system with $\partial \phi / \partial t = -\nabla \cdot \mathbf{J} = 0$, subject to the impermeability conditions for $\mathbf{v}$ and $\mathbf{j}$, and the no-slip condition for $\mathbf{v}$ at the solid surface. The corresponding total stress fields and total fluxes are $\left( \boldsymbol{\sigma}^{(1)}, \mathbf{J}^{(1)} \right)$ and $\left( \boldsymbol{\sigma}^{(2)}, \mathbf{J}^{(2)} \right)$, respectively. Now we assume that the stationary states are close to equilibrium, with $\mathbf{v}$, $\phi - \phi_{\text{eq}}$, $\nabla p$, $\nabla \cdot \boldsymbol{\sigma}$, $\nabla \mu$, $\mathbf{j}$, and $\mathbf{J}$ regarded as leading order deviation from equilibrium. Using the stationary state condition $\nabla \cdot \mathbf{J} = 0$, the definition equation $\mathbf{J} = \phi \mathbf{v} + \mathbf{j}$, the force balance equation (10), the constitutive equation (11) for $\mathbf{j}$, the definition equation (12) $\boldsymbol{\sigma} = -p\mathbf{I} + \boldsymbol{\sigma}_{\text{visc}}$, the constitutive equation (13) for $\boldsymbol{\sigma}_{\text{visc}}$, and the incompressibility condition $\nabla \cdot \mathbf{v} = 0$, we obtain

$$\begin{aligned} &-\int_{\partial\Omega} \mu^{(1)} \mathbf{n} \cdot \mathbf{J}^{(2)} dA + \int_{\partial\Omega} \mathbf{n} \cdot \boldsymbol{\sigma}^{(1)} \cdot \mathbf{v}^{(2)} dA \\ &= \int_{\Omega} M^{-1} \mathbf{j}^{(1)} \cdot \mathbf{j}^{(2)} d\mathbf{r} + \int_{\Omega} \frac{\eta}{2} \left[ \nabla \mathbf{v}^{(1)} + \left( \nabla \mathbf{v}^{(1)} \right)^T \right] : \left[ \nabla \mathbf{v}^{(2)} + \left( \nabla \mathbf{v}^{(2)} \right)^T \right] d\mathbf{r} \end{aligned}, \tag{A22}$$

up to the quadratic order in $\mathbf{j}$ and $\nabla \mathbf{v}$, which measure the deviation from equilibrium. Here the phenomenological coefficients $M$ and $\eta$ are treated as equilibrium properties independent of the dynamic state. This is for stationary states close to equilibrium. (In general, the shear viscosity $\eta$ and the mobility $M$ may depend on the local volume fraction $\phi$, i.e. $\eta = \eta(\phi)$ and $M = M(\phi)$. However, if the stationary state is close to equilibrium, then we have $\eta \approx \eta(\phi_{eq})$ and $M \approx M(\phi_{eq})$ in Eq. (A22).) As a result, Eq. (A22) becomes permutable for the two stationary states, leading to the GLRT in the form of

$$-\int_{\partial\Omega} \mu^{(1)} \mathbf{n} \cdot \mathbf{J}^{(2)} dA + \int_{\partial\Omega} \mathbf{n} \cdot \boldsymbol{\sigma}^{(1)} \cdot \mathbf{v}^{(2)} dA = -\int_{\partial\Omega} \mu^{(2)} \mathbf{n} \cdot \mathbf{J}^{(1)} dA + \int_{\partial\Omega} \mathbf{n} \cdot \boldsymbol{\sigma}^{(2)} \cdot \mathbf{v}^{(1)} dA. \quad (A23)$$

Furthermore, using $\mu_{eq} = \text{const.}$, $\boldsymbol{\sigma}_{eq} = -p_{eq}\mathbf{I}$, and $p_{eq} = \text{const.}$, we rewrite the above equation as

$$-\int_{IO} \left(\mu^{(1)} - \mu_{eq}\right) \mathbf{n} \cdot \mathbf{J}^{(2)} dA + \int_{IO} \mathbf{n} \cdot \left(\boldsymbol{\sigma}^{(1)} - \boldsymbol{\sigma}_{eq}\right) \cdot \mathbf{v}^{(2)} dA = -\int_{IO} \left(\mu^{(2)} - \mu_{eq}\right) \mathbf{n} \cdot \mathbf{J}^{(1)} dA + \int_{IO} \mathbf{n} \cdot \left(\boldsymbol{\sigma}^{(2)} - \boldsymbol{\sigma}_{eq}\right) \cdot \mathbf{v}^{(1)} dA$$

$$, \quad (A24)$$

in which $\mu - \mu_{eq}$ and $\mathbf{n} \cdot (\boldsymbol{\sigma} - \boldsymbol{\sigma}_{eq})$ are regarded as the generalized forces due to deviation from the equilibrium state. The GLRT in Eq. (A24) can be readily rewritten in the general form of the GLRT in Eq. (24), $\left[\mathbf{F}^{(1)}\right]^T \mathbf{I}^{(2)} = \left[\mathbf{F}^{(2)}\right]^T \mathbf{I}^{(1)}$, as presented in Sec. IIC.

## 3. Micropolar fluids

The classical LRT can also be generalized for the study of the slow flows of micropolar fluids [10], which represent one of the most well understood complex fluids with microstructures. Consider a micropolar fluid in a volume region $\Omega$ with a solid boundary denoted by $\partial\Omega$. Suppose there are two stationary state solutions $\left(\mathbf{v}^{(1)}, \boldsymbol{\omega}^{(1)}\right)$ and $\left(\mathbf{v}^{(2)}, \boldsymbol{\omega}^{(2)}\right)$ for the dynamic system with $\partial \rho / \partial t = -\nabla \cdot (\rho \mathbf{v}) = 0$, subject to the impermeability condition for $\mathbf{v}$ and the no-slip conditions for $\mathbf{v}$ and $\boldsymbol{\omega}$ at the solid surface. The corresponding total stress and couple-stress fields are $\left(\boldsymbol{\sigma}^{(1)}, \mathbf{C}^{(1)}\right)$ and $\left(\boldsymbol{\sigma}^{(2)}, \mathbf{C}^{(2)}\right)$, respectively. Now we assume that the stationary states are close to equilibrium, with $\rho \approx \rho_{eq}$ and $\nabla \cdot \mathbf{v} \approx 0$. Using these two equations, the balance equations (39) and (40), and the constitutive equations (41), (42) and (43), we can obtain

$$\int_{\partial\Omega} \mathbf{n}\cdot\boldsymbol{\sigma}^{(1)}\cdot\mathbf{v}^{(2)}dA + \int_{\partial\Omega} \mathbf{n}\cdot\mathbf{C}^{(1)}\cdot\boldsymbol{\omega}^{(2)}dA =$$

$$\int_{\Omega}\left[\kappa(\nabla\cdot\mathbf{v}^{(1)})(\nabla\cdot\mathbf{v}^{(2)}) + 2\eta\mathbf{E}_v^{(1)}:\mathbf{E}_v^{(2)}\right]d\mathbf{r} + \int_{\Omega}\left[\nu_1(\nabla\cdot\boldsymbol{\omega}^{(1)})(\nabla\cdot\boldsymbol{\omega}^{(2)}) + 2\nu_2\mathbf{E}_\omega^{(1)}:\mathbf{E}_\omega^{(2)}\right]d\mathbf{r}, \quad (A25)$$

$$+\int_{\Omega}\xi\left(\frac{1}{2}\nabla\times\mathbf{v}^{(1)} - \boldsymbol{\omega}^{(1)}\right)\left(\frac{1}{2}\nabla\times\mathbf{v}^{(2)} - \boldsymbol{\omega}^{(2)}\right)d\mathbf{r}$$

up to the quadratic order in the rates that measure the deviation from equilibrium. Here the five phenomenological coefficients on the right-hand side of equation are treated as equilibrium properties independent of the dynamic state. This is for stationary states close to equilibrium. (In general, these coefficients may depend on the local density $\rho$. However, if the stationary state is close to equilibrium, then we have $\rho \approx \rho_{eq}$ and these coefficients become equilibrium properties.) As a result, Eq. (A25) becomes permutable for the two stationary states, leading to the GLRT in the form of

$$\int_{\partial\Omega}\mathbf{n}\cdot\boldsymbol{\sigma}^{(1)}\cdot\mathbf{v}^{(2)}dA + \int_{\partial\Omega}\mathbf{n}\cdot\mathbf{C}^{(1)}\cdot\boldsymbol{\omega}^{(2)}dA = \int_{\partial\Omega}\mathbf{n}\cdot\boldsymbol{\sigma}^{(2)}\cdot\mathbf{v}^{(1)}dA + \int_{\partial\Omega}\mathbf{n}\cdot\mathbf{C}^{(2)}\cdot\boldsymbol{\omega}^{(1)}dA. \quad (A26)$$

Furthermore, using $\boldsymbol{\sigma}_{eq} = -p_{eq}\mathbf{I}$, $p_{eq} = \text{const.}$, and $\mathbf{C}_{eq} = 0$, we rewrite the above equation as

$$\int_{\partial\Omega}\mathbf{n}\cdot(\boldsymbol{\sigma}^{(1)} - \boldsymbol{\sigma}_{eq})\cdot\mathbf{v}^{(2)}dA + \int_{\partial\Omega}\mathbf{n}\cdot(\mathbf{C}^{(1)} - \mathbf{C}_{eq})\cdot\boldsymbol{\omega}^{(2)}dA = \int_{\partial\Omega}\mathbf{n}\cdot(\boldsymbol{\sigma}^{(2)} - \boldsymbol{\sigma}_{eq})\cdot\mathbf{v}^{(1)}dA + \int_{\partial\Omega}\mathbf{n}\cdot(\mathbf{C}^{(2)} - \mathbf{C}_{eq})\cdot\boldsymbol{\omega}^{(1)}dA$$
$$, \quad (A27)$$

in which $\mathbf{n}\cdot(\boldsymbol{\sigma} - \boldsymbol{\sigma}_{eq})$ and $\mathbf{n}\cdot(\mathbf{C} - \mathbf{C}_{eq})$ are regarded as the generalized forces due to deviation from the equilibrium state. The GLRT in Eq. (A27) can be readily rewritten in the general form of the GLRT in Eq. (24), $\left[\mathbf{F}^{(1)}\right]^T \mathbf{I}^{(2)} = \left[\mathbf{F}^{(2)}\right]^T \mathbf{I}^{(1)}$, as presented in Sec. IIC.

### 4. Thermal conduction in solids and still fluids

Finally, we generalize the LRT for thermal conduction in solids and still fluids that can be inhomogeneous and anisotropic [8, 23, 24]. Consider a solid in a volume region $\Omega$ with a boundary denoted by $\partial\Omega$. Suppose there are two temperature fields $T^{(1)}$ and $T^{(2)}$ which are stationary state solutions of the thermal conduction equation, i.e. $\nabla\cdot\mathbf{q}^{(1)} = \nabla\cdot\mathbf{q}^{(2)} = 0$, with $\mathbf{q}^{(1)}$ and $\mathbf{q}^{(2)}$ being the heat current density corresponding to $T^{(1)}$ and $T^{(2)}$, respectively. (Note that heat transfer between the solid and its surrounding environment is necessary for the system to maintain nontrivial stationary states with $\mathbf{q} \neq 0$.) Using the constitutive equation (60) for $\mathbf{q}$, we can obtain

$$\int_{\partial\Omega} \frac{1}{T^{(1)}} \mathbf{n} \cdot \mathbf{q}^{(2)} dA = \int_{\Omega} \mathbf{q}^{(2)} \cdot \nabla \frac{1}{T^{(1)}} d\mathbf{r} = \int_{\Omega} \mathbf{q}^{(2)} \cdot \boldsymbol{\lambda}^{-1}(T^{(1)}) \cdot \mathbf{q}^{(1)} d\mathbf{r} , \qquad (A28)$$

in which the stationary state condition $\nabla \cdot \mathbf{q} = 0$ has been used as well. For stationary states in the immediate proximity of equilibrium state at which $T = T_{eq} = \text{const.}$, we can use $\boldsymbol{\lambda}(T^{(1)}) \approx \boldsymbol{\lambda}(T^{(2)}) \approx \boldsymbol{\lambda}(T_{eq})$ in Eq. (A28) up to the quadratic order in $\mathbf{q}$. As a result, Eq. (A28) becomes permutable for the two stationary states, leading to the GLRT in the form of

$$\int_{\partial\Omega} \frac{1}{T^{(1)}} \mathbf{n} \cdot \mathbf{q}^{(2)} dA = \int_{\partial\Omega} \frac{1}{T^{(2)}} \mathbf{n} \cdot \mathbf{q}^{(1)} dA . \qquad (A29)$$

Furthermore, using $T_{eq} = \text{const.}$ and $\int_{\partial\Omega} \mathbf{n} \cdot \mathbf{q} dA = 0$ from $\nabla \cdot \mathbf{q} = 0$, we rewrite the above equation as

$$\int_{\partial\Omega} \left( \frac{1}{T^{(1)}} - \frac{1}{T_{eq}} \right) \mathbf{n} \cdot \mathbf{q}^{(2)} dA = \int_{\partial\Omega} \left( \frac{1}{T^{(2)}} - \frac{1}{T_{eq}} \right) \mathbf{n} \cdot \mathbf{q}^{(1)} dA . \qquad (A30)$$

in which $1/T - 1/T_{eq}$ is regarded as a thermodynamic force due to deviation from the equilibrium state. The GLRT in Eq. (A30) can be readily rewritten in the general form of the GLRT in Eq. (24), $\left[ \mathbf{F}^{(1)} \right]^T \mathbf{I}^{(2)} = \left[ \mathbf{F}^{(2)} \right]^T \mathbf{I}^{(1)}$, as presented in Sec. IIC.